\documentclass{article}
\usepackage{spconf,amsmath,graphicx}
\usepackage[shortlabels]{enumitem}
\DeclareMathOperator*{\argmax}{argmax}
\usepackage{algorithm}
\usepackage{algpseudocode}
\usepackage{hyperref}
\usepackage{multirow}
\usepackage[belowskip=0pt,aboveskip=5pt]{caption}

\def\L{{\cal L}}
\def\A{{\cal A}}

\def\S{{\cal S}}

\title{ONLINE ACTIVE LEARNING FOR SOUND EVENT DETECTION}
\twoauthors
 {Mark Lindsey, Ankit Shah, Richard M. Stern\thanks{This work is supported by a corporate gift from Probity, Inc.}}
	{Carnegie Mellon University\\
	5000 Forbes Ave, Pittsburgh, PA 15213}
 {Francis Kubala}
	{Probity, Inc.\\
	754 Elden St, Herndon, VA 20170}
\begin{document}
\ninept
\maketitle
\begin{abstract}
Data collection and annotation is a laborious, time-consuming prerequisite for supervised machine learning tasks. Online Active Learning (OAL) is a paradigm that addresses this issue by simultaneously minimizing the amount of annotation required to train a classifier and adapting to changes in the data over the duration of the data collection process. Prior work has indicated that fluctuating class distributions and data drift are still common problems for OAL. This work presents new loss functions that address these challenges when OAL is applied to Sound Event Detection (SED). Experimental results from the SONYC dataset and two Voice-Type Discrimination (VTD) corpora indicate that OAL can reduce the time and effort required to train SED classifiers by a factor of 5 for SONYC, and that the new methods presented here successfully resolve issues present in existing OAL methods.
\end{abstract}
\begin{keywords}
Active learning, online learning, sound event detection, data drift
\end{keywords}
\section{Introduction}
\label{sec:intro}

Data annotation has always been a bottleneck for supervised training of machine learning models. This issue is particularly prevalent for tasks like Sound Event Detection (SED), which can be cognitively taxing to annotate. The annotation requirement also prohibits potential practitioners from using the model while data collection is still in process. Even after enough data has been collected to achieve reasonable base performance, additional data annotation is required to adapt the model to specific environments.

\begin{figure}[htb]
    \centering
    \includegraphics[scale=0.43]{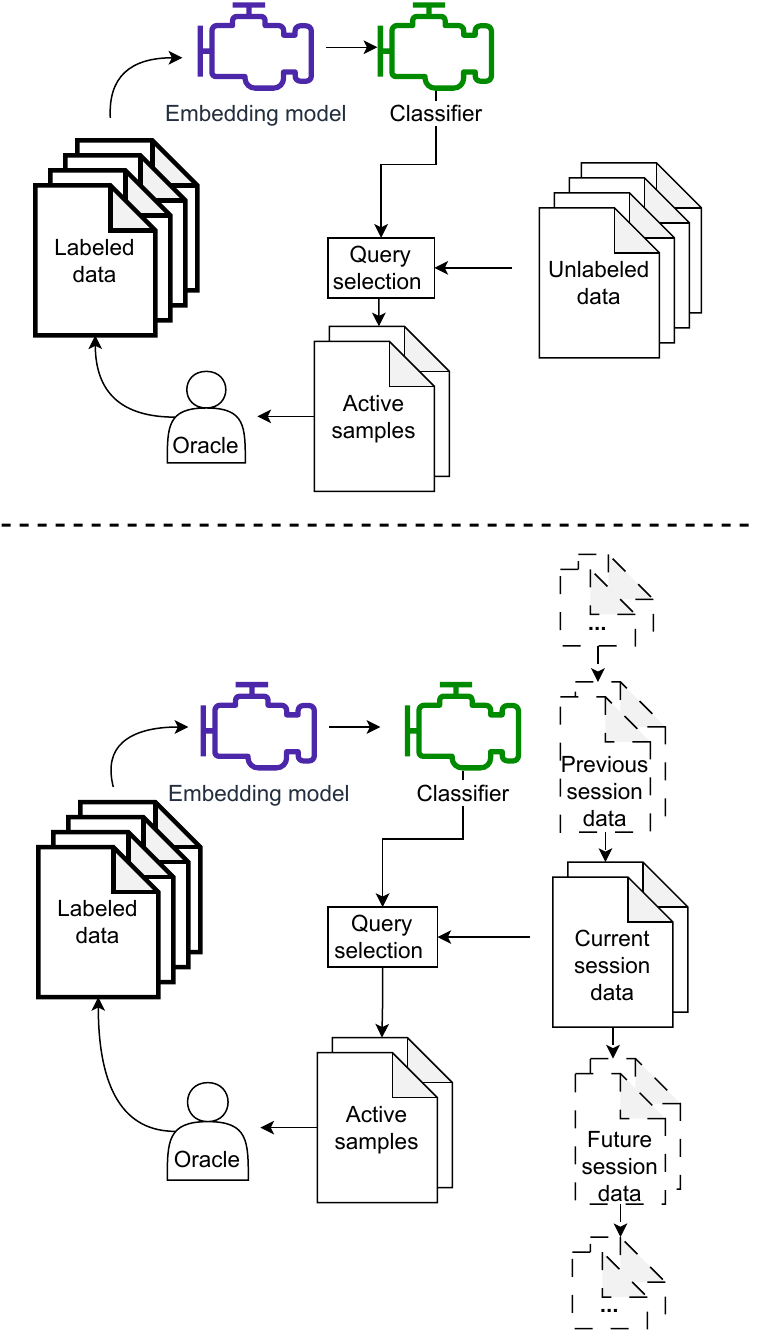}
    \caption{\centering AL (top) vs. OAL (bottom). For OAL, the current session is updated every step.}
    \label{fig:al_vs_oal}
\end{figure}

Various alternative paradigms have been developed to address the issues posed by data annotation. These paradigms include self-supervised training, unsupervised methods, and active learning. Active Learning (AL) reduces the number of annotations required to train a model by employing an algorithm that actively identifies data points that would be the most informative if the label were known \cite{al_survey}. While AL can reduce the absolute amount of annotation required for training, typical AL approaches are not designed to start training before data collection is complete or to facilitate adaptation. However, a subfield of AL, referred to as Online Active Learning (OAL), takes AL a step further by adding an online learning component (see Fig. \ref{fig:al_vs_oal}). This addition allows training to begin before all the data has been collected. Thus, OAL serves as a method to reduce even more of the data annotation bottleneck than AL.

The addition of online learning poses new challenges that do not exist for AL. Possibly the most significant of these problems is handling data drift over time. Data drift in online learning scenarios requires the classifier to adapt in order to maintain reasonable performance. For OAL, the query selection strategy must also be aware of or robust to data drift, as query selection is crucial to classifier adaptation. 

Data drift also poses a particular challenge for detection tasks where it is critical to avoid missed detections. Many of such detection tasks use the Detection Cost Function (DCF) as the evaluation metric. DCF is a weighted combination of False Negative Rate (FNR) and False Positive Rate (FPR) where FN errors are more costly than FP errors (three times greater in this work). Typical loss functions, like cross-entropy loss, attempt to optimize overall classification accuracy regardless of the types of errors the classifier makes. Such loss functions do not automatically take into account class imbalance or weighted error types and must be manually adjusted based on a prior knowledge of the problem.

This work introduces an OAL training scheme that specifically seeks to reduce the required annotation for SED. This work also presents new loss functions intended to handle varying class distributions in all paradigms, including OAL, to optimize the DCF. SED experiments are performed on the SONYC dataset and two Voice-Type Discrimination (VTD) corpora, showing that OAL can both reduce the required annotations by a factor of 5 and start training much earlier while achieving comparable DCF. Experiments performed with the new loss functions show that it can reduce the DCF and FNR for fully-supervised and AL training by up to 30\% relative to the cross-entropy loss results. However, the same pattern does not hold when the loss functions are used in OAL training.

\section{Related Work}
\label{sec:review}
Work on OAL is relatively sparse but is gaining traction. Much of the research for batch-based OAL is focused on using drift detection algorithms \cite{drift_review} to determine when to make adjustments to the model or the AL training parameters \cite{oal_survey}. These adjustments include increasing or decreasing AL query density \cite{adjust_budget}, weighting long-term and short-term models \cite{dynamic_weight}, or introducing completely new models \cite{new_model} based on perceived data drift. Despite the temporal nature of audio, most published OAL methods are not applied to audio tasks, likely, in part, because of a lack of online audio datasets. The authors believe that this work is the first to apply OAL to SED.

Methods for training with imbalanced data are represented well in the literature. Examples include focal loss \cite{focal_loss}, which down-weights samples that are easily classified in training, and losses that use dynamic reweighting based on known or estimated class distributions \cite{dynamic_imbalance, balance_loss}. Some AL methods also address class imbalance because of the strong effect it can have on performance \cite{al_imbalance}. All of these approaches differ from the loss functions presented here, which directly optimize the DCF regardless of class distribution.

\section{Methodology}
\label{sec:method}
This paper includes two primary methodological contributions: 1) the application of OAL to SED, and 2) loss functions that optimize the DCF for imbalanced class distributions.

\subsection{OAL for SED}
\label{sec:oal_method}
Typical datasets are not structured in a way that is usable for OAL training. This section describes how to convert pre-existing datasets with temporal information into OAL datasets, as well as the basic steps to apply OAL to the SED task.

Datasets with spatial markers should be organized into samples from the same \textit{environment}. In the case of audio data, an environment refers to a sensor in a fixed location. Within each environment, data should be put in chronological order based on time of occurrence. From there, the samples can be grouped into \textit{sessions} (or batches) of $L$ abutting samples. Additionally, a \textit{bootstrap corpus} is formed to initialize the classifier. A bootstrap corpus of size $N$ is composed of the first $N/2$ occurrences of each class.

With the data organized in this manner, Algorithm \ref{algo:oal} can be applied to simultaneously train a classifier and make predictions about the data from each environment using OAL. These steps are also illustrated and contrasted with regular AL \cite{aloe} in Fig. \ref{fig:al_vs_oal}.

\begin{algorithm} 
\caption{Online Active Learning for SED} \label{algo:oal}
    \begin{algorithmic}
    \Require classifier $\Theta$, query selection strategy $\Phi$, bootstrap corpus $C$, OAL sessions $\S$, query budget $B$, adaptation data pool $\A$
    \\
    \State Initialize $\Theta$ with all samples from $C$
    \State Add all samples from $C$ to $\A$
    \While{not all sessions in $\S$ have been seen}
    \State Load new session $\S_i$ from $\S$
    \State Run $\Phi$ on $\S_i$ to select $B$ most informative data $X_j$
    \State Obtain the labels $y_j$ for the queried samples
    \State Add the labeled data  $(X_j, y_j)$ to $\A$
    \State Update $\Theta$ with all samples from $\A$
    \State Predict $\hat{y}_j$ for the unlabeled samples in $\S_i$
    \EndWhile
    \end{algorithmic}
\end{algorithm}

\subsection{DCF-based Loss Functions}
New loss functions based on DCF are presented here to address the challenges of class imbalance and weighted error metrics, which are common in OAL. The error rates that define DCF are typically calculated using a non-differentiable argmax operation, so the traditional formulation of DCF itself is not a viable neural network loss function. The following approximations for FNR and FPR were made to ensure that the DCF is differentiable and directly optimizable by a neural network. Here, $y_i$ is the label (0 or 1) and $\hat{y}_i$ is the posterior probability of sample $i$ being part of the target class.

\begin{equation}\label{eq:fn}
    \hat{p}_{fn} = \frac{\sum_i (1-\hat{y}_i)\cdot y_i}{\sum_i y_i}
\end{equation}

\begin{equation}\label{eq:fp}
    \hat{p}_{fp} = \frac{\sum_i \hat{y}_i\cdot (1-y_i)}{\sum_i (1-y_i)}
\end{equation}


Eqs. \ref{eq:fn} and \ref{eq:fp} replace the argmax function with expectations to calculate the error rates in a differentiable manner. As such, this loss function is referred to as the expected DCF (e-DCF) loss. To closer approximate the true DCF, a differentiable argmax function (d-argmax) can be used \cite{dargmax}. D-argmax is implemented as shown in Eq. \ref{eq:lim} with Softmax function $\sigma$ and multiplier $\lambda$. Here, $x$ refers to a class index and $f(x)$ is the posterior probability that a given sample belongs to that class. In the limit, the term $\sigma(\lambda f(x))$ reduces to 0 if $x$ is not the most likely class, or 1 if it is the most likely.

\begin{equation} \label{eq:lim}
    \text{d-}\argmax_x f(x) = \lim_{\lambda\to\infty}  \sum_x  x \sigma(\lambda f(x)) \approx \sum_x x \sigma(\lambda f(x))
\end{equation}

For the purpose of implementation, a large-valued $\lambda$ can be used in place of the infinite limit as an approximation. With this adjustment, the $\hat{y}_i$ terms in Eqs. \ref{eq:fn} and \ref{eq:fp} can be replaced with $\sigma(\lambda \hat{y}_i)$, i.e., the Softmax of the posterior probability of the target class scaled by $\lambda$. Compared to $\hat{y}_i$, this term will be much closer to 1 or 0, depending on the predicted class. This implementation of the DCF loss is referred to as differentiable DCF (d-DCF).

\section{Experiments}
\label{sec:exp}

\subsection{Datasets}
All experiments were performed on data from the SONYC Urban Sound Tagging (SONYC-UST) dataset \cite{sonyc} and the SRI and Lionbridge (LB) VTD corpora.

\subsubsection{SONYC Dataset}
The SONYC dataset\footnote{\href{https://zenodo.org/record/3966543}{https://zenodo.org/record/3966543}} consists of 18,515 audio clips from 61 microphones placed in different locations around New York City. A predefined split reserves 13,538 of these samples for training, 4,308 for validation, and 669 for testing. Each clip is 10 seconds in duration and recorded with a sample rate of 48 kHz. Annotations for eight coarse-grained classes and 23 fine-grained classes are provided for each sample. Only the coarse-grained classes are considered in this work. Information regarding the time and place of recording and sensor ID is also provided as part of the annotations.

\subsubsection{VTD Corpora}
VTD is the task of detecting live speech, i.e., speech that is produced spontaneously within the recording environment. As such, VTD can be viewed as SED with a single target class---live speech.

The SRI Corpus comprises 1,617 hours of speech, recorded in 4 different rooms by microphones in 5 locations in each room. One microphone from each room was used in this work. Individual recordings range from 3.5 to 14 hours in duration. 11.8\% of the audio is target audio, and the distractors consist of TV, radio, traffic, room noises, and pre-recorded audio from the Linguistic Data Consortium. The LB Corpus contains 2,966 hours of audio, recorded in 3 rooms by 7 microphones in each room. Again, only one microphone from each room was used. Individual recording durations range from 1.8 to 8.3 hours. Target live speech is present in 23.8\% of the audio. Distractors include TV, podcasts, and ambient noise.

Annotations for these two corpora are provided as start and stop timestamps for target class audio. For the purposes of this work, the annotations are reduced to a binary indication of the presence of target audio for every abutting five-second audio frame. Both VTD corpora will be released publicly in the coming months.

\subsection{Classifier Architecture and Hyperparameters}

\begin{figure}
    \centering
    \includegraphics[scale=0.4]{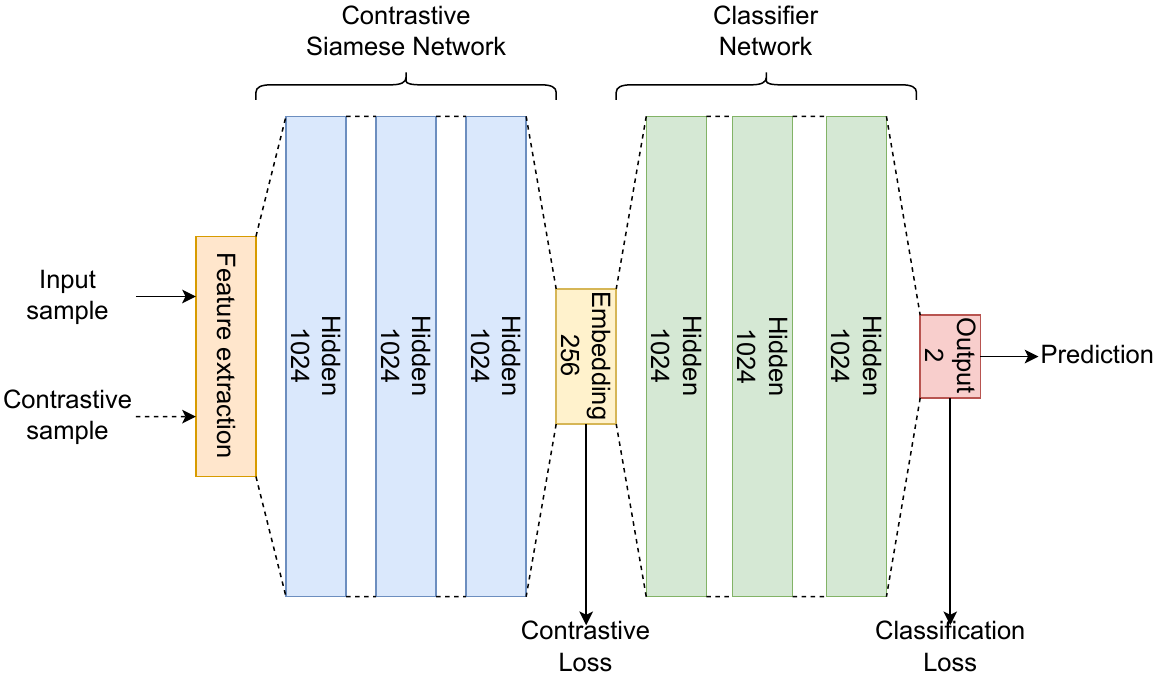}
    \caption{The contrastive classifier architecture.}
    \label{fig:arch}
\end{figure}

The neural network used for all experiments is the contrastive classifier depicted in Fig. \ref{fig:arch}. As indicated, it is trained with a combined contrastive loss and classification loss. The only difference between the classifier used on the SONYC data and the VTD data is the input features. The input features for the SONYC data are either Wav2CLIP\footnote{\href{https://github.com/descriptinc/lyrebird-wav2clip}{https://github.com/descriptinc/lyrebird-wav2clip}} or CLAP\footnote{\href{https://huggingface.co/laion/clap-htsat-fused}{https://huggingface.co/laion/clap-htsat-fused}} embeddings (512 dimensions each) \cite{clap1, clap2, wav2clip} or a concatenation of the two. In comparison, the VTD data is represented by WavLM\footnote{\href{https://huggingface.co/microsoft/wavlm-large}{https://huggingface.co/microsoft/wavlm-large}} embeddings (1024 dimensions) \cite{wavlm}, so the classifier used for VTD is larger in most cases.

All other hyperparameters used during training are the same across experiments. The learning rate is fixed at $10^{-4}$ using an Adam optimizer, weight decay is set at $10^{-5}$, and early stopping after five epochs of no improvement. The contrastive loss \cite{contrastive_loss} is described by Eq. \ref{eq:contrastive_loss_fn}, where $x_1$ and $x_2$ are the embedding vectors of the input samples being contrasted, $y$ denotes whether the samples come the same class (1) or different classes (0), $d$ is the Euclidean distance between the sample embeddings, and $m$ is the margin (set to 1 here). The total loss is equal to the average of the contrastive loss and the classification loss.

\begin{equation} \label{eq:contrastive_loss_fn}
    \L_c(\mathbf{x_1}, \mathbf{x_2}, y) = y \cdot d^2 + (1-y)\cdot \max (m-d,0)^2
\end{equation}


\subsection{Preliminary Experiments}
Before testing the new methods presented in this work, the basic experimental setup was compared to existing methods to ensure that performance was in a reasonable range. The systems submitted to the DCASE2020 Challenge \cite{DCASE2020Workshop} serve as benchmarks. The evaluation metric used for these experiments is the Area Under the Precision-Recall Curve (AUPRC), since this is the metric reported by the challenge. The default data splits are also retained to make a viable comparison to the challenge baseline. The performance of the contrastive classifier was evaluated using Wav2CLIP, CLAP, and both embeddings concatenated as input features. Only fully-supervised training was used for these experiments. The input features with the best overall AUPRC are used in the subsequent experiments.

\subsection{OAL Experiments}
\label{sec:oal_exp}
The OAL experiments are run as explained in Section \ref{sec:oal_method} and compared to fully supervised training. Experiments are performed on the SONYC dataset. The fully-supervised experiments are performed in the same way as described for the preliminary experiments. For the OAL experiments, an environment was defined as a single sensor that contributes 10 minutes or more audio to the dataset. This criterion results in 47 valid environments in the dataset. A session in each of these environments was defined as 30 consecutive ten-second samples or 5 minutes of audio. In each session, a budget of 5 labeled samples was allowed to be queried for classifier training, and the bootstrap corpus for each environment was defined to contain 8 samples total. The query strategy of choice was uncertainty sampling based on negative energy \cite{negenergy, wang2022active}. The target class for these experiments is human speech.

These experiments are intended to show that the OAL setup can achieve similar performance as fully-supervised training. These experiments are also intended to show the extent to which required data collection and annotation can be reduced. The DCF is used to evaluate the performance of the systems in these experiments. Note that the DCFs will not be exactly comparable across supervised and OAL experiments since the OAL experiments do not use the same test split as the other experiments. However, the DCF should still provide a rough estimate of model performance. The reduction in required data is measured in two ways: 1) the number of labeled samples used to train the classifier and 2) the number of samples required to be in the collection before training can begin.

\subsection{Loss Function Experiments}
To evaluate the DCF-based loss functions, the contrastive classifier was trained with a cross-entropy loss and both DCF loss functions using fully-supervised training, AL, and OAL. The DCF loss was compared to unweighted cross-entropy, and cross-entropy with a weighting ratio of 4:1 for the target class compared to the non-target class. The fully-supervised experiments were performed on the default data splits of the SONYC dataset. The AL experiments also used the same test splits, but the training set was adjusted to be used as the unlabeled data pool from which samples were queried every AL step. The AL experiments basically followed the same procedure as illustrated in the top panel of Fig. \ref{fig:al_vs_oal}. All supervised experiments and AL experiments were repeated 5 times. The OAL experiments were run one time on both datasets in the same manner as outlined in Section \ref{sec:oal_exp}. All loss function experiments are evaluated in terms of DCF, FNR, and FPR. All AL and OAL experiments were done with the negative energy-based query strategy. Again, the target class for the experiments using SONYC data is human speech, which is present in 36\% of the dataset. 

\section{Results and Discussion}
\label{sec:results}
\subsection{Preliminary Results}


\begin{table}[tbh]
    \centering
    \begin{tabular}{c|c}
        Classifier & AUPRC $\uparrow$ \\
        \hline
        DCASE2020 baseline & 0.5100 \\
        \hline
        Wav2CLIP & 0.4147 \\
        CLAP & \textbf{0.5208} \\
        Wav2CLIP+CLAP & 0.5158 \\
    \end{tabular}
    \caption{\centering Macro-AUPRC comparison for fully-supervised training}
    \label{tab:prelim}
\end{table}

The results of the contrastive classifier with the three sets of input features is compared to the challenge baseline in Table \ref{tab:prelim}. Note that these numbers represent the average AUPRC across all 8 coarse-grained classes. This number is referred to as ``Macro-AUPRC" in the DCASE2020 Challenge. It is clear from the table that the best-performing system is the contrastive classifier with CLAP embeddings. In light of these results, CLAP embeddings are used as the input feature of choice for the OAL and loss function experiments.


\subsection{OAL Results}

Table \ref{tab:oal_results} compares the results of OAL and fully-supervised training for SONYC. While fully-supervised training shows better performance across all error-related metrics, OAL training achieves competitive results with far fewer labeled samples. This suggests that OAL could be a practical approach for reducing the required annotations, even if it comes at the cost of somewhat higher error rates. In this case, OAL used only 3,291 samples, which equates to 18\% of the training and validation set used for fully-supervised training.

\begin{table}[tbh]
    \centering
    \begin{tabular}{c|c|c}
        Metric & Fully-supervised & OAL \\
        \hline
        DCF & 0.1714 & 0.2117 \\
        FNR & 0.1661 & 0.1983 \\
        FPR & 0.1871 & 0.2517 \\
        \# labels & 17,846 & 3,291 \\
        \# samples to start & 18,515 & 30 \\
    \end{tabular}
    \caption{\centering Comparing full supervision to OAL using SONYC data}
    \label{tab:oal_results}
\end{table}

The most significant advantage of OAL is made evident by the last metric in the table. For fully-supervised training, all 18,515 samples comprising the training, validation, and test sets must be collected, and all training and validation samples must be labeled before classifier training begins. This starkly contrasts the mere 30 unlabeled samples required to start training the same classifier with OAL.

Results on the VTD data are also encouraging. Using only 40 seconds of labeled audio per hour (i.e., 1.1\% of the data), a DCF of 0.0718 is achieved (see Table \ref{tab:oal_dcf_loss}). Note that model training can begin after only one hour of collected data.

\subsection{Loss Function Results}

\begin{table}[tbh]
    \centering
    \begin{tabular}{c|c|c|c|c}
         Paradigm & Loss Fn & DCF $\downarrow$ & FNR $\downarrow$ & FPR $\downarrow$ \\
         \hline
         \multirow{4}{*}{Supervised} & XENT (1:1) & 0.1946 & 0.2169 & \textbf{0.1277} \\
          & XENT (4:1) & 0.1714 & 0.1661 & 0.1871 \\
          & e-DCF & \textbf{0.1467} & 0.1329 & 0.1883 \\
          & d-DCF & 0.1557 & \textbf{0.1153} & 0.2768 \\
        \hline
        \multirow{4}{*}{AL} & XENT (1:1) & 0.2077 & 0.2332 & 0.1311 \\
          & XENT (4:1) & 0.1999 & 0.2280 & \textbf{0.1154} \\
          & e-DCF & 0.1720 & 0.1805 & 0.1468 \\
          & d-DCF & \textbf{0.1602} & \textbf{0.1681} & 0.1367 \\
    \end{tabular}
    \caption{\centering Error rates for unweighted and weighted cross-entropy (XENT), e-DCF, and d-DCF losses for fully-supervised and AL experiments on the SONYC dataset, averaged over five runs}
    \label{tab:dcf_loss}
\end{table}

\begin{table}[tbh]
    \centering
    \begin{tabular}{c|c|c|c|c}
         Dataset & Loss Fn & DCF $\downarrow$ & FNR $\downarrow$ & FPR $\downarrow$ \\
        \hline
               & XENT (4:1) & \textbf{0.2117} & \textbf{0.1983} & \textbf{0.2517} \\
         SONYC & e-DCF & 0.2129 & 0.1996 & 0.2530 \\
               & d-DCF & 0.3126 & 0.3018 & 0.3451 \\
        \hline
               & XENT (4:1) & \textbf{0.0718} & \textbf{0.0884} & \textbf{0.0219} \\
           VTD & e-DCF & 0.0934 & 0.1159 & 0.0260 \\
               & d-DCF & 0.1166 & 0.1423 & 0.0397 \\
    \end{tabular}
    \caption{\centering Error rates for weighted cross-entropy (XENT), e-DCF, and d-DCF losses for OAL experiments on the SONYC and VTD datasets}
    \label{tab:oal_dcf_loss}
\end{table}

Table \ref{tab:dcf_loss} shows the advantage of DCF-based loss functions over cross-entropy for fully supervised and AL training. This advantage can be seen most prominently in terms of FNR, which was reduced by 30.6\% for full supervision and 26.3\% for AL when d-DCF loss replaces cross-entropy. Such an improvement in FNR over the weighted cross-entropy indicates that DCF-based loss functions do indeed optimize DCF well, even when classes are imbalanced.

However, Table \ref{tab:oal_dcf_loss} shows that DCF-based loss functions do not have the same advantage in all cases. For OAL on either SONYC or VTD data, cross-entropy is better than the DCF loss functions in every error metric. It is unclear why this is the case, but it may be a consequence of dealing with very small amounts of data in the under-represented SONYC environments.

\section{Conclusions}
This work addresses the problem of reducing the cost of data annotation for SED by training classifiers using OAL. New loss functions intended to handle class imbalance and weighted error metrics like DCF are also introduced and evaluated. Experiments on the SONYC dataset show that OAL can effectively reduce the number of annotations required by a factor of 5 and allow training to begin after collecting only 30 samples instead of the whole dataset. OAL for the VTD dataset can begin after one hour of data is collected and achieves a DCF of 0.0718 while only requiring labels from 1.1\% of the whole dataset. The DCF-inspired loss functions yield major reductions in DCF (up to 20\% relative) and FNR (up to 30\% relative) for fully-supervised and AL training. Future work might include making improvements to the OAL setup or developing loss functions that improve performance for OAL.


\bibliographystyle{IEEEbib}
\bibliography{refs}

\begin{thebibliography}{10}

\bibitem{al_survey}
Burr Settles,
\newblock ``{Active Learning Literature Survey},''
\newblock 2009.

\bibitem{drift_review}
Jie Lu, Anjin Liu, Fan Dong, Feng Gu, João Gama, and Guangquan Zhang,
\newblock ``{Learning under Concept Drift: A Review},''
\newblock {\em IEEE Transactions on Knowledge and Data Engineering}, vol. 31,
  no. 12, pp. 2346--2363, 2019.

\bibitem{oal_survey}
Davide Cacciarelli and Murat Kulahci,
\newblock ``{A Survey on Online Active Learning},''
\newblock {\em arXiv preprint arXiv:2302.08893}, 2023.

\bibitem{adjust_budget}
Andrea Castellani, Sebastian Schmitt, and Barbara Hammer,
\newblock ``{Stream-Based Active Learning with Verification Latency in
  Non-stationary Environments},''
\newblock in {\em International Conference on Artificial Neural Networks}.
  Springer, 2022, pp. 260--272.

\bibitem{dynamic_weight}
Hang Zhang, Weike Liu, and Qingbao Liu,
\newblock ``{Reinforcement Online Active Learning Ensemble for Drifting
  Imbalanced Data Streams},''
\newblock {\em IEEE Transactions on Knowledge and Data Engineering}, vol. 34,
  no. 8, pp. 3971--3983, 2022.

\bibitem{new_model}
Bartosz Krawczyk, Bernhard Pfahringer, and Michał Woźniak,
\newblock ``Combining active learning with concept drift detection for data
  stream mining,''
\newblock in {\em 2018 IEEE International Conference on Big Data (Big Data)},
  2018, pp. 2239--2244.

\bibitem{focal_loss}
{Lin, Tsung-Yi and Goyal, Priya and Girshick, Ross and He, Kaiming and Dollar,
  Piotr},
\newblock ``{Focal Loss for Dense Object Detection},''
\newblock in {\em Proceedings of the IEEE International Conference on Computer
  Vision (ICCV)}, Oct 2017.

\bibitem{dynamic_imbalance}
{Fernando, K. Ruwani M. and Tsokos, Chris P.},
\newblock ``{Dynamically Weighted Balanced Loss: Class Imbalanced Learning and
  Confidence Calibration of Deep Neural Networks},''
\newblock {\em {IEEE Transactions on Neural Networks and Learning Systems}},
  vol. 33, no. 7, pp. 2940--2951, 2022.

\bibitem{balance_loss}
Yin Cui, Menglin Jia, Tsung-Yi Lin, Yang Song, and Serge Belongie,
\newblock ``{Class-Balanced Loss Based on Effective Number of Samples},''
\newblock in {\em Proceedings of the IEEE/CVF Conference on Computer Vision and
  Pattern Recognition (CVPR)}, June 2019.

\bibitem{al_imbalance}
{Yu, Hualong and Yang, Xibei and Zheng, Shang and Sun, Changyin},
\newblock ``{Active Learning From Imbalanced Data: A Solution of Online
  Weighted Extreme Learning Machine},''
\newblock {\em {IEEE Transactions on Neural Networks and Learning Systems}},
  vol. 30, no. 4, pp. 1088--1103, 2019.

\bibitem{aloe}
Harlin Lee, Aaqib Saeed, and Andrea~L Bertozzi,
\newblock ``{Active Learning of Non-Semantic Speech Tasks with Pretrained
  Models},''
\newblock in {\em ICASSP 2023-2023 IEEE International Conference on Acoustics,
  Speech and Signal Processing (ICASSP)}. IEEE, 2023, pp. 1--5.

\bibitem{dargmax}
David Diaz-Guerra, Antonio Miguel, and Jose~R. Beltran,
\newblock ``{Direction of Arrival Estimation of Sound Sources Using Icosahedral
  CNNs},''
\newblock {\em IEEE/ACM Transactions on Audio, Speech, and Language
  Processing}, vol. 31, pp. 313--321, 2023.

\bibitem{sonyc}
Mark Cartwright, Ana Elisa~Mendez Mendez, Jason Cramer, Vincent Lostanlen,
  Graham Dove, Ho-Hsiang Wu, Justin Salamon, Oded Nov, and Juan Bello,
\newblock ``{SONYC urban sound tagging (SONYC-UST): A multilabel dataset from
  an urban acoustic sensor network},''
\newblock 2019.

\bibitem{clap1}
Yusong Wu, Ke~Chen, Tianyu Zhang, Yuchen Hui, Taylor Berg-Kirkpatrick, and
  Shlomo Dubnov,
\newblock ``{Large-scale Contrastive Language-Audio Pretraining with Feature
  Fusion and Keyword-to-Caption Augmentation},''
\newblock in {\em IEEE International Conference on Acoustics, Speech and Signal
  Processing, ICASSP}, 2023.

\bibitem{clap2}
Ke~Chen, Xingjian Du, Bilei Zhu, Zejun Ma, Taylor Berg-Kirkpatrick, and Shlomo
  Dubnov,
\newblock ``{HTS-AT: A Hierarchical Token-Semantic Audio Transformer for Sound
  Classification and Detection},''
\newblock in {\em IEEE International Conference on Acoustics, Speech and Signal
  Processing, ICASSP}, 2022.

\bibitem{wav2clip}
Ho-Hsiang Wu, Prem Seetharaman, Kundan Kumar, and Juan~Pablo Bello,
\newblock ``{Wav2CLIP: Learning Robust Audio Representations from CLIP},''
\newblock in {\em ICASSP 2022-2022 IEEE International Conference on Acoustics,
  Speech and Signal Processing (ICASSP)}. IEEE, 2022, pp. 4563--4567.

\bibitem{wavlm}
Sanyuan Chen, Chengyi Wang, Zhengyang Chen, Yu~Wu, Shujie Liu, Zhuo Chen, Jinyu
  Li, Naoyuki Kanda, Takuya Yoshioka, Xiong Xiao, et~al.,
\newblock ``{WavLM: Large-scale Self-supervised Pre-training for Full Stack
  Speech Processing},''
\newblock {\em IEEE Journal of Selected Topics in Signal Processing}, vol. 16,
  no. 6, pp. 1505--1518, 2022.

\bibitem{contrastive_loss}
S.~Chopra, R.~Hadsell, and Y.~LeCun,
\newblock ``{Learning a Similarity Metric Discriminatively, with Application to
  Face Verification},''
\newblock in {\em {2005 IEEE Computer Society Conference on Computer Vision and
  Pattern Recognition (CVPR'05)}}, 2005, vol.~1, pp. 539--546 vol. 1.

\bibitem{DCASE2020Workshop}
Nobutaka Ono, Noboru Harada, Yohei Kawaguchi, Annamaria Mesaros, Keisuke Imoto,
  Yuma Koizumi, and Tatsuya Komatsu,
\newblock {\em {Proceedings of the Fifth Workshop on Detection and
  Classification of Acoustic Scenes and Events (DCASE 2020)}},
\newblock Tokyo, Japan, November 2020.

\bibitem{negenergy}
Weitang Liu, Xiaoyun Wang, John Owens, and Yixuan Li,
\newblock ``{Energy-based Out-of-distribution Detection},''
\newblock {\em Advances in neural information processing systems}, vol. 33, pp.
  21464--21475, 2020.

\bibitem{wang2022active}
Xin Wang and Junichi Yamagishi,
\newblock ``{Investigating Active-Learning-Based Training Data Selection for
  Speech Spoofing Countermeasure},''
\newblock in {\em 2022 IEEE Spoken Language Technology Workshop (SLT)}, 2023,
  pp. 585--592.

\end{thebibliography}

\end{document}